\documentclass[journal,compsoc]{IEEEtran} 

\usepackage{makeidx}
\usepackage{graphicx}
\usepackage{multirow}
\usepackage{multicol}
\usepackage{rotating}
\usepackage{cite}
\usepackage{caption}
\usepackage{subcaption}

\usepackage{amssymb}
\usepackage{epsfig}
\usepackage{epstopdf}
\usepackage{float}
\usepackage{booktabs} 
\usepackage{url}
\usepackage{amsmath} 
\usepackage{enumitem} 
\usepackage{mathtools}
\usepackage{tikz}

\usepackage[left=0.6in,top=0.6in,right=0.6in,bottom=0.6in]{geometry}
\usepackage[ruled,vlined]{algorithm2e}
\usepackage{algpseudocode}
\usepackage{pifont}
\usepackage{tikz}
\usepackage{multirow} 
\usepackage{hhline}
\ifCLASSINFOpdf
\usepackage{soul}
\else
\fi

\begin{document}

\title{A Secure and Trusted Mechanism for Industrial IoT Network using Blockchain}

\author{Geetanjali~Rathee, Farhan Ahmad,~\IEEEmembership{Member,~IEEE}, Naveen Jaglan, and \\ 
Charalambos Konstantinou,~\IEEEmembership{Senior Member,~IEEE}
	\IEEEcompsocitemizethanks{
	
	\IEEEcompsocthanksitem (Corresponding Authors: Geetanjali Rathee and Farhan Ahmad)
	\IEEEcompsocthanksitem G. Rathee is with the Department of Computer Science and Engineering, Netaji Subhas University of Technology, Dwarka Sector-3, New Delhi, India. (e-mail: geetanjali.rathee123@gmail.com)\protect
	
	\IEEEcompsocthanksitem F. Ahmad is with the Systems Security Group, Centre for Future Transport and Cities, Coventry University,  Warwickshire CV1 5FB, United Kingdom (e-mail: ad5899@coventry.ac.uk)\protect
	\IEEEcompsocthanksitem N. Jaglan is with the Department of Electronics and Communication Engineering, Jaypee University of Technology, Waknaghat-173234, India. (e-mail: naveenjaglan1@gmail.com)\protect
    \IEEEcompsocthanksitem Charalambos Konstantinou is with the Computer, Electrical and Mathematical Sciences and Engineering (CEMSE) Division, King Abdullah University of Science and Technology (KAUST), Thuwal 23955-6900, Saudi Arabia. (e-mail: charalambos.konstantinou@kaust.edu.sa)\protect
}
}


\IEEEtitleabstractindextext{
\begin{abstract} \setlength{\leftskip}{0pt} \setlength{\rightskip}{0pt}
Industrial Internet-of-Things (IIoT) is a powerful IoT application which remodels the growth of industries by ensuring transparent communication among various entities such as hubs, manufacturing places and packaging units. Introducing data science techniques within the IIoT improves the ability to analyze the collected data in a more efficient manner, which current IIoT architectures lack due to their distributed nature. From a security perspective, network anomalies/attackers pose high security risk in IIoT. In this paper, we have addressed this problem, where a coordinator IoT device is elected to compute the trust of IoT devices to prevent the malicious devices to be part of network. Further, the transparency of the data is ensured by integrating a blockchain-based data model. The performance of the proposed framework is validated extensively and rigorously via MATLAB against various security metrics such as attack strength, message alteration, and probability of false authentication. The simulation results suggest that the proposed solution increases IIoT network security by efficiently detecting malicious attacks in the network.
\end{abstract}

\begin{IEEEkeywords}
Industrial Internet-of-Things (IIoT), Blockchain, Security, Secure IoT Devices, Trust Management.
\end{IEEEkeywords}
}

\maketitle

\IEEEdisplaynontitleabstractindextext
\IEEEpeerreviewmaketitle



\section{Introduction}
\label{sec:Introduction}


Today the performance and productivity of an organization entirely depends on the way it analyses and collects its business data. The onset of smart systems along with various other developments in the field of data science continue to provide new frontiers for the expansion of this technology. According to the recent statistics, the number of devices with online connectivity stand at 6 billion which collectively generate approximately 2.5 Quintilian bytes of data \cite{karpatne2017theory}.

In conventional scenarios, collecting and analysing static data from devices in real-time was inefficient. Today, these devices communicate with each other, thanks to the novel Internet-of-Things (IoT) paradigm, resulting in generation of information without any human intervention \cite{Farhan_Commag}. Further, intelligent and smart objects/sensors working mutually are steadily becoming more expansive and venture to accomplish users' demands. In order to extract and address meaningful information from the data, a new field, namely, data science that uses scientific approaches, algorithms, procedures and systems for analysis and collection of huge data has been proposed recently. Data science in IoT (DS-IoT) is a technique that improves the online collection and analysis of information in a more scientific, realistic and efficient way\cite{oh2019personal}. DS-IoT integrates a diverse range of smart devices with commercial objects that export manufacturing information through sensors worn in the field of medicine, cyber physical systems and transportation and to retain the records \cite{merolla2014million}.

Nowadays, DS-IoT is considered as an important technique within industries to increases their growth, effectiveness and overall efficiency \cite{wang2014cellular, bertino2017botnets}. Further, a wide range of IoT-based applications such as smart cities, e-healthcare, intelligent transportation and Industrial IoT (IIoT) have been pioneer to aid intelligent verdict making by concerning a series of physical objects vital to the experimental escalation in an efficient and effective manner \cite{zhou2017computation}. Amongst a variety of use cases offered by IoT technology, IIoT is considered as an important application of IoT that controls and traces every activity of the industry for its growth \cite{huang2019towards}. IIoT refers to the network where data is collected from numerous sensors, actuators, and machines within an industrial environment and is accessed through the Internet \cite{al2018context, 8621042}. 

		 \begin{figure}[ht]
 	    	 \centering
  	    	 \includegraphics[width=0.45\textwidth]{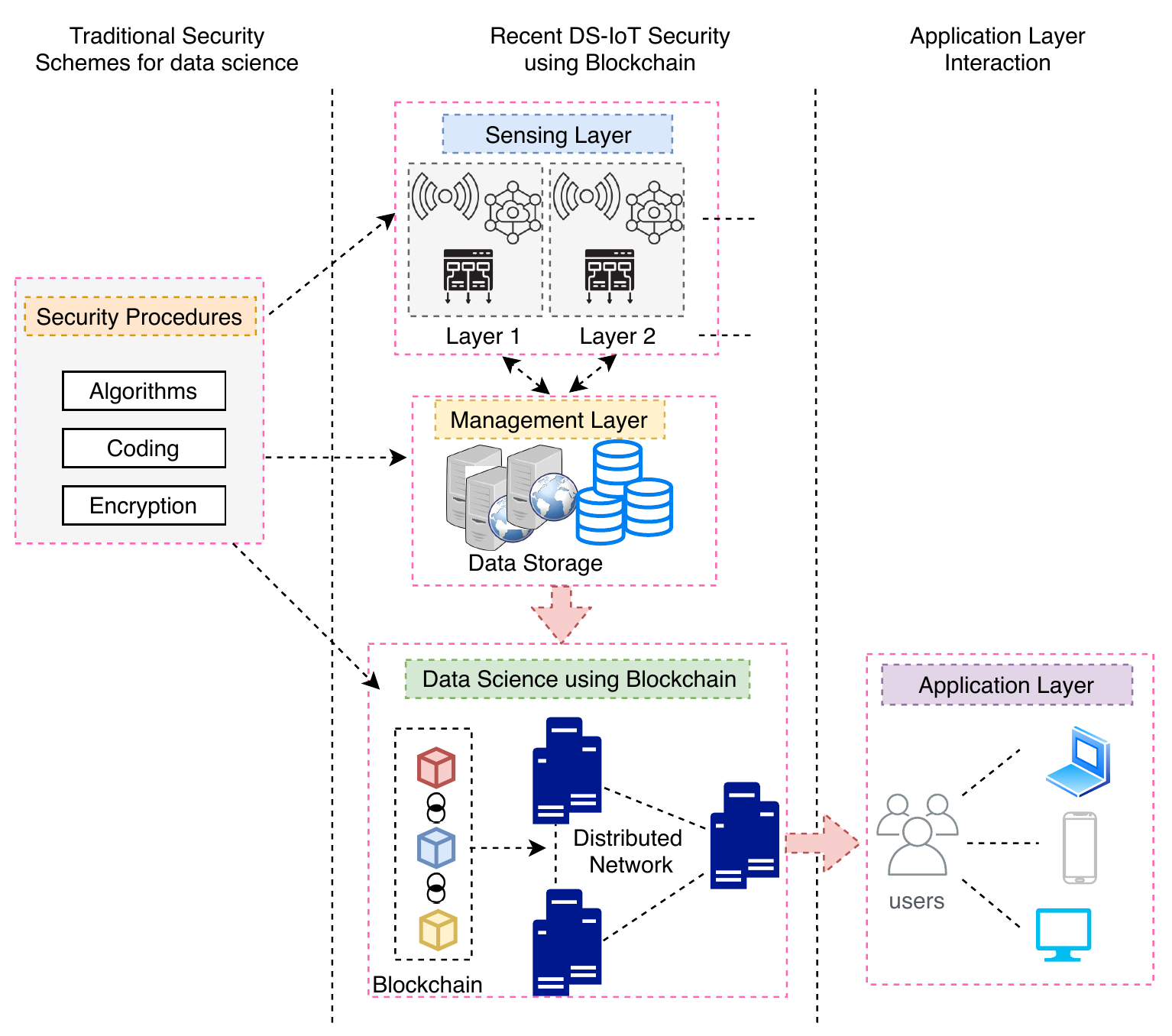}
  	    	 \caption{DS-IoT Blockchain in Industrial IoT}
  	    	 \label{fig:DS-IoT}
		 \end{figure}

\subsection{Motivation}
Despite of various advantages DS-IoT technique offers, organizations are still reluctant to use IoT devices due to several security concerns \cite{shamsi2018understanding}. From the security perspective, a malicious device within the industry premises can degrade the network performance by limiting the legitimate IoT devices from sharing trusted and true information or alter the communication data. Even though a few open-source ciphers are still liable to bugs and exploits, they are persistently scrutinised by numerous users and turn less effective to malevolent modifications from centralised entities or third parties \cite{li2019enhancing}. In conventional industrial applications, all the smart devices are often assumed to be cooperative and trusted. However, in practice and reality, IoT devices are prone to mischievous activities of the malicious devices (MD). Thus, the potential challenge is to distinguish the ideal DS-IoT devices from the malicious ones in order to establish a legitimate communication environment \cite{moosavi2016delay}. In addition, transparency plays an imperative role in increasing trust and security among the authorities so that any alteration in data can be immediately recognized by the owners \cite{chen2014collaborative}. Moreover, to prevent future modifications of data captured by smart devices, recently, blockchain has been proposed, where a network maintains a chain of blocks consisting of the data recorded by the IoT devices in IIoT environment while manufacturing and shipping the products as depicted in Fig. \ref{fig:DS-IoT}.  

Blockchain provides an efficient and transparent mechanism for analyzing and controlling the data. Any alteration in the data by any user can be identified via Blockchain. In order for concurrent handling of events for brisk responses and secluded monitoring, the evolution of IIoT generates the possibility of connecting automated systems. Further, data science approaches also ensure effective data gathering and processing mechanisms and techniques for IIoTs. Although DS-IoT in industries has various benefits, many organisations and businesses are still reluctant to use it. The exploitation of this IoT scheme is still steep due to the high cost of allied centralized clouds and servers \cite{danzi2019delay, liu2018evolutionary, sun2019blockchain}. To the best of author’s knowledge, the DS-IoT security through blockchain is unexploited for IIoT in the reported literature.
IoT devices in the IIoT environment transmit very sensitive information, such as the temperature of the boiler, product manufacturing record, document or product shipping records etc. Therefore, it must be a paramount necessity that security by design is provided to the IIoT network. Further the presence of malicious nodes may have severe impact on the IIoT network as they have the ability to tamper the data generated by IoT devices. Therefore, this leads to the question about how to provide a secure networking paradigm for enabling the devices to share data in an attack-free environment.

\subsection{Contributions of the paper}
In order to analyze real-time data in IIoT while manufacturing and shipping the products, we have introduced an orchestration of security concerns with the aim to detect and resolve the threats caused by an intruder. To do so, we have computed the trust factor (TF) of all the devices using an elected Coordinator IoT Device (CID). The CID is the controlling unit of IoT environment that is responsible for verifying the legitimacy of the IoT devices present in the network. Further, in order to prevent future alteration of stored information in the local database, a  private blockchain is introduced within IIoT environment to keep track of all the recorded information stored in the database. Therefore, the potential contributions of this paper are as follows:

\begin{itemize}
    \item A novel security orchestration for IIoT using trust to gather and maintain huge amount of data generated by IoT devices has been proposed.
    
    \item A mathematical model has been derived to identify the probability of error generated in IIoT by exploiting the concept of probability of false IoT authentication and non-detection.
    
    \item A novel data security model using blockchain in IIoT to prevent the intruders from altering the stored information in a local database has been proposed.
    
    \item The performance of the proposed framework has been rigorously evaluated against probabilistic hypothetical scenarios for both small and large IIoT networks using a simulation model.
\end{itemize}

The rest of the paper is structured as follows. Section \ref{sec:2} is dedicated to the related work in IIoT and the use of blockchain in IIoT. Section \ref{sec:3} provides details of our proposed solution using blockchain to secure IIoT networks. Next, Section \ref{sec:4} provides the details of the simulation model and the results. Lastly, Section \ref{sec:5} concludes and directs the prospect of the paper. 

\section{Related Work}
\label{sec:2}
Various studies have been carried out to secure IIoT networks using data science and blockchain. For instance, Yan et al. \cite{yan2017industrial} explored the issues of data processing for industrial big data by proposing novel structural multi source information for heterogeneous environments. This framework is validated by analyzing the heterogeneous data of industries. Even though the authors have discussed the use of smart devices for data driven mechanisms in industries by focusing on storage, processing and utilization schemes, they did not consider the ways through which stored data can be maintained and can further be compromised by various intruders. Further, Wang et al. \cite{wang2019new} proposed a novel industrial data processing mechanism to apprehend various industrial functions including distributed access, storage, stream and batch data processing, and real time controlling. Compared to traditional data processing schemes, \cite{wang2019new} have illustrated various features of analyzing, correlating and integrating huge amount of data in IIoT. However, the generation of such huge data may further lead to various complexity concerns such as data storage, communication through intermediate nodes and transmission cost. Therefore, secure data transmission through legitimate intermediate nodes is still a lingering question for researchers. 

Recently, the most promising technique which adds decentralization, trust, privacy and security to diverse IIoT fields is blockchain. To ensure secure information delivery via IoT devices, S. Yu et al. \cite{yu2018high} proposed a blockchain-based mechanism to transmit the data with minimum cost and economic transfer value. Various techniques such as distributed network architecture, consent algorithm and mapping of intelligent devices were used to identify the decentralized autonomy of smart devices. Furthermore, Y. Yu et al. \cite{yu2018blockchain} addressed the issue of security and privacy concerns in IoT objects and proposed a blockchain-IoT framework. Authenticated scalability, decentralized schemes and assertion of data transfer for the payments are the several facilities offered with the blockchain facilitated IoT infrastructure. Further, the proposed phenomenon is validated by illustrating certain solutions using Ethereum by presenting the embedding of blockchain within IoT. However, the type of blockchain (public/private) through which intruders may further compromise intermediate IoT devices is not mentioned in this study.  

Oh et al. \cite{oh2020competitive} have proposed a data trading mechanism to ensure the privacy among business stakeholders for IoT markets. The authors have used Nash equilibrium to measure the feasibility and maximized the profits of market stakeholders. Hasan et al.\cite{hasan2022trustworthy}  have proposed an interplanetary file system mechanism while streaming and storing the data generated by IoT devices. The authors have ensured the security of the proposed mechanism using blockchain by generating the smart contracts, algorithms and diagrams with their complete implementation process. The authors have showed the novelty and effectiveness of the proposed system as compare to traditional scheme. Lam et al. \cite{lam2022dynamical} have proposed a decentralized automatic orchestration and configuration mechanism based upon semantic policies. The proposed approach was deployed and verified while sending the information during planning and production in IIoT environment while transmitting over cloud. 


Though various studies have proposed a wide range of techniques to ensure a decentralized, transparent and secure mechanism in IIoT networks, very few of them pointed the number of attacking strategies of the intruders aiming to disrupt or consume the network resources. Further, none of the authors relied on trust-based mechanisms to detect the nodes' legitimacy, data storage or processing techniques through blockchain specifically for IIoT networks.In summary, data science techniques within IoT were focused on various studies due to the advantages it offer as mentioned earlier. Further, few studies also focused on blockchain to provide security in IIoT. However, introducing both data science and blockchain within IIoT network can increase network efficiency in terms of efficient industrial data analytic in a secure environment. Therefore, this study provides a novel and secure framework for IIoT by integrating both data science and blockchain to identify possible threats within the network. In the next section, we provide details of our proposed framework. 

\section{Proposed Industrial Blockchain Framework}
\label{sec:3}
\begin{figure}[ht]
   	 \centering
 \includegraphics[width=0.45\textwidth]{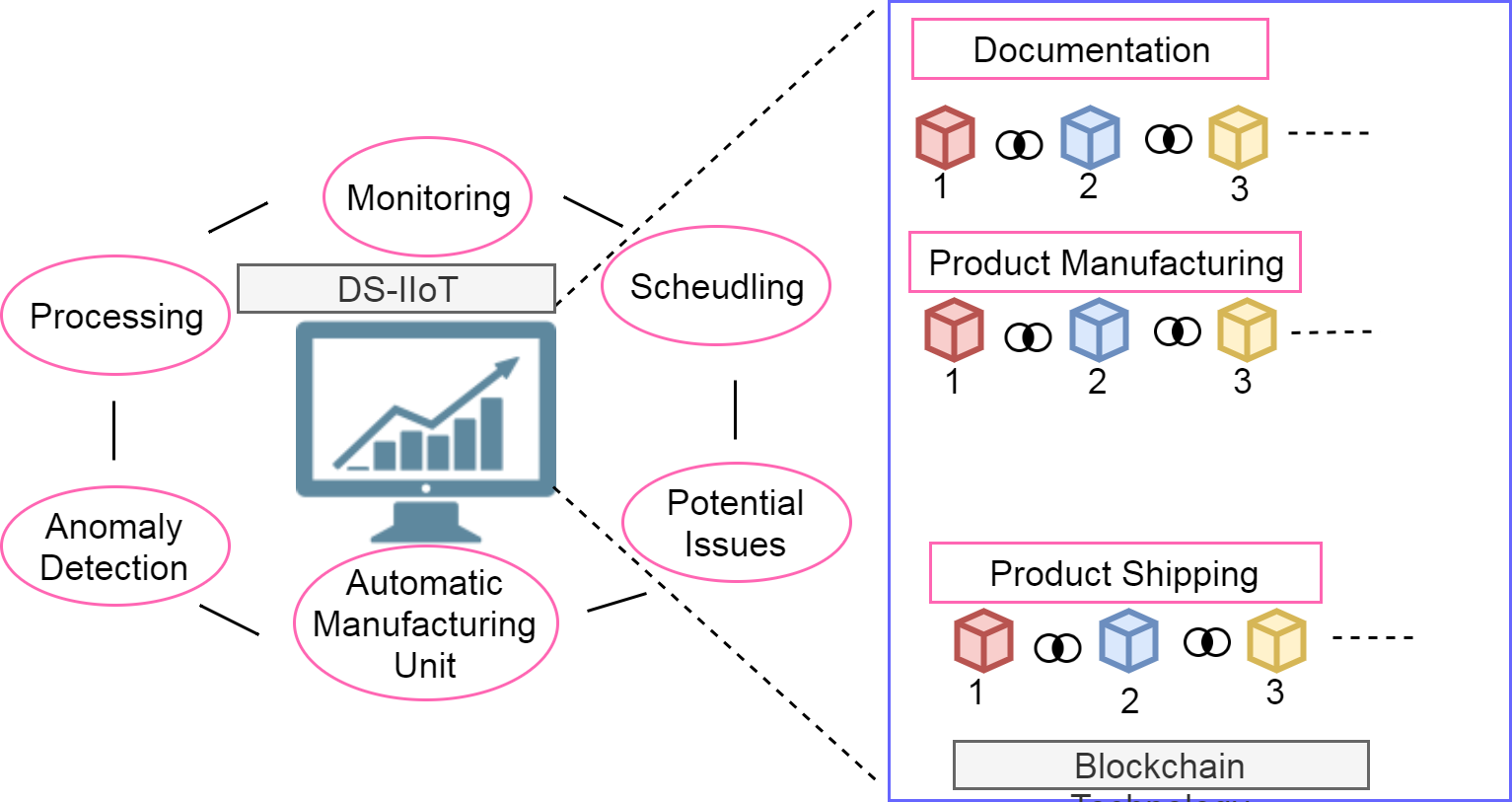}
   	    	 \caption{Data Model of DS-IIoT using Blockchain}
   	    	 \label{fig:Data_Model_in_IIoT}
 		 \end{figure}

In order to describe our proposed solution, we considered an IIoT network, including both legitimate and malicious devices. Next, a system model is developed to validate the proposed framework. Further, a blockchain is integrated within the IIoT network to ensure secure data analysis in the proposed DS-IIoT environment. Finally, a mathematical model is derived to evaluate the performance of the network. The blockchain technique is able to build control systems and data sharing system for CU in order to address the challenges of decentralized information circulation, internal information controlling access and privacy while sharing the data among various entities. Figure \ref{fig:Data_Model_in_IIoT} depicts the data sharing mechanism through blockchain where each record of every individual entity is stored on a blockchain that can be further traced and analyzed by all the users. The malicious data record or alteration in stored information can be immediately identified by all the entities though blockchain technique. 

Further, Figure \ref{fig:Data_Model_in_IIoT} highlights the blockchain-based data model in IIoT networks, suggesting that various phases such as generation, manufacturing, processing, monitoring, and anomaly detection in data can easily be resolved by maintaining separate  blockchain of every phase. Moreover, by scheduling every transaction on blockchain further enhances the overall processing capabilities by maintaining a transparent relationship among the IoT devices. 


\subsection{System Model}
Traditionally, the role of IoT devices in IIoT networks has been to monitor or control the product manufacturing, documentation, record and share it with the back-end server. In case of an attack, the intruders may breach the security of the network to steal confidential data such as industry records, employees' information and business statistical data which might affect the overall trade of the industry. Hence, in practice, IIoT networks  are susceptible to attackers who tend to perform malicious activities in the IoT environment. Further, malicious devices behave intelligently in a random manner to remain undetected in order to bypass the monitoring phase of the network. Furthermore, the stored information about workers' location, activity processing, in-out time and product manufacturing record may also be stolen by the intruders for their personal usage. To tackle these sensitive issues, a novel framework has been introduced in this study, which integrates a trust-based solution and blockchain within IIoT.

The system model of the proposed framework consists of `$n$' IoT devices which are assumed to be trusted initially. Among $n$ IoT devices, the responsible to verify the legitimacy of the IoT devices is elected on the basis of two criteria, i.e., (1) energy level and (2) monitoring capability (MC). The energy level indicates that the IoT device has sufficient energy to monitor and coordinate the activities and communication between neighbouring devices. The monitoring capability of the device, on the other hand, identifies the malicious behavior within the network. In this proposal, any IoT device with high energy level and monitoring capability is chosen as CID. Energy of the IoT devices and CID is calculated via Equation \ref{eq: Energy}.

 	\begin{equation}
	E_{i}=\sum_{p=1}^{n} \mid {X_{(i,p)}}\mid^{2} = \\
	\begin{cases}
	E_i \geq \gamma \ presence \ of \ IoT \ device,\\
	E_i \leq \gamma\  absence \ of \ IoT \ device.\\
	
	\end{cases}   
	\label{eq: Energy}
	\end{equation}

Where X${_{(i,p)}}$ is the $p^{th}$ sample of the $i^{th}$ IoT device, $\gamma$ is a predefined threshold value to monitor and $n$ is the total number of IoT devices. Further, MC of IoT devices is computed as Equation \ref{eq: MC}.

\begin{equation}
    MC_{i}=NDG_{i}=\sum_{i=1}^{n} (PDR_i + E_i + Activeness_i) \\
    \label{eq: MC}
\end{equation}

Where $MC_i$ is measured in terms of $NDG_i$ (Network Development Goal) that detects the status and health of the network. The node's status and health is categorized into three colors, i.e., (1) \textit{black} where nodes are identified as malevolent, (2) \textit{green} for legitimate nodes, and (3) \textit{grey} to indicate alert messages upon entry of a new node or extensive transmission of existing node in the network. DG is further dependent upon certain parameters. Packet Delivery Ratio ($PDR_i$) defines the number of packets forwarded by a particular node. $E_i$ is the energy consumed by a node to transmit or forward the incoming packets and is calculated using Equation \ref{eq: Energy}. Finally $Activeness_i$ shows how much time a node remains active in the network.

The MC of the CID further depends on the survival time (ST) of the device within the IIoT network. The device having sufficient energy level along with oldest survival time (OST) would be elected as the coordinator of radio environment consisting of table including CID identity, IoT device identity, CID address, ID address, TF and ST of each device. The oldest survival time of IoT device is defined as \textit{the total time period of the IoT device to remain alive in the network}. The CID computes the trust factor of each IoT device by using various characteristics such as monitoring capability, communication among the users and actions of all the devices.

Depending on the nature of the IoT device, the potential response of CID falls into following two categories:

\subsubsection{CASE 1: When existing IoT device is identified as MD}

Whenever, the IoT device starts behaving maliciously, then two possibilities exist, i.e., either the IoT device is MD or it is disrupted. In both the cases, a random IoT device is forged and compromised by the intruders. In our proposed system, initial device is identified via Equation \ref{eq:device}.

\begin{equation}
\label{eq:device}
	IoT \ Device =
	\begin{cases}
	 LID: & \text{$ST_{ID}$ \textgreater $ST_{MD}$ with transmission $T$}\\
	 MD: & \text{$ST_{ID}$ \textless $ST_{MD}$ no transmission $T$}\\
	\end{cases}       
	\end{equation}

Where, LID and MD are legitimate and malicious devices respectively. If the ST of the IoT device is higher than that of MD, then the device is identified as legitimate, otherwise, it is assumed to be compromised. Next, the TF of each IoT device is computed if the device is either compromised or transmits very large number of messages in the network. To achieve this, various parameters including previous history, MC and communication behavior (CB) are taken into account. CB is defined specially for malicious IoT devices which transmit huge number of false messages towards legitimate devices. MC and CB of any node satisfying predefined thresholds, TF is considered as 1. Otherwise, it is 0. 

\subsubsection{CASE 2: When new IoT device is identified as MD}

In the case where a new IoT device (NID) is identified in the environment, the possibility of cases can be raised as either legitimate or malicious. In both scenarios, as the ST of NID is very less as compared to that of existing IoT devices, therefore, initially CID allows at least five communication transmissions to the NID. TF of a legitimate NID is 1, and for malicious NID, TF would always be below the predefined threshold range. Each node is assigned an initial trust value ranging from 0.7- 0.95 (as assumed) with 1 as the highest trust value. This trust is increased or decreased by checking the social rank of each node by using predefined threshold values. The main reason behind this value is that the threshold will be around 30\% of the maximum value, so taking initial value range will help in the further calculation instead of taking every value as simply 1. There is a disadvantage of taking this scenario as few nodes tend to lose the trust early as compared to the other nodes. CID keeps all the records and information of the IoT devices in its table. After specified number of transmissions, TF of the NID would be checked by CID and corresponding action would be taken according to the Equation \ref{eq:NID}.

\begin{equation}
\label{eq:NID}
	NID =
	\begin{cases}
	TF==1, & \text{then legitimate ID,}\\
	TF==0, & \text{then malicious device.}\\
	
	\end{cases}       
	\end{equation}

If TF of NID is 1, then ID is identified as trusted and allowed further transmissions, else all of its further transmissions would be blocked by considering it as an MD. 

\subsection{Alteration of stored information}
The presence of the intruders with the intention to hack or compromise the network cannot be ignored as IIoT networks contain very sensitive information. Therefore, the proposed solution integrates blockchain in the back-end to ensure transparency in the network. The stored records are kept on blockchain so that any alteration or deletion of any information inside the IIoT network can be traced easily. The Algorithm \ref{algo:main_algo} represents complete execution of the proposed framework. Further, Algorithms \ref{algo:TF} and \ref{algo:MC} show the high level algorithms to calculate the resulting functions TF() and MC() within the main algorithm.

\begin{algorithm}[h]
\SetAlgoLined
\textbf{Assumption:} $Count_{threshold}$ = 50\%\\
 \textbf{Input:} (1) Network with `n' IDs, (2) Among them one CID is elected, and (3) `m' number of MD's\\
\textbf{Output:} ID identified as either legitimate or malicious \\
The CID selection is based on ST, energy level and MC. \\
CID maintains a table having ID id, ID address, routing information, CID id, ST and TF of each ID to identify MD.
Upon the emergence, the NID is identified as MD else legitimate. \\
  \eIf{(ID  is NID)}{
       CID allows first five assumptions and\;
       Compute TF()\;
       Compute MC()\;
       Blockchain record ()\; 
       The information of each record corresponding to ID is stored in the database with its current and previous hashes.
    }{
    ID is elected as MD
    }
\caption{Execution of Proposed IIoT Framework}
\label{algo:main_algo}
\end{algorithm}

\begin{algorithm}[b]
\SetAlgoLined
 \textbf{Input:} The number of transactions/communications done by each IoT device are recorded to compute their TF \\
  \eIf{(ID’s previous history and MC() satisfies predefined threshold value )}{
      	Set TF=1 to ID\;
      	return 1\;
       }{
   Set TF=0\;
   Mark ID as MD\;
   return 0\;
    }
\caption{Calculation of TF()}
\label{algo:TF}
\end{algorithm}

\begin{algorithm}[h]
\SetAlgoLined
 \textbf{Input:} The number of interactions or communications among IoT devices \\
  \eIf{(ID trace wrong information and store incorrect data )}{
      	Set C = C+1 \;
        \eIf{(Set C $>$ $C_{threshold}$)}{ID is MD\; return 0\;}{IoT device is legitimate and trusted}
       }{IoT device is malicious}
\caption{Calculation of MC()}
\label{algo:MC}
\end{algorithm}

\subsection{Mathematical Model of the Proposed Approach }
In order to validate the proposed framework, MDs are deployed randomly in the IIoT network with the following two aims: (1) to forge the identity of a legitimate device, and (2) to consume the network resources by generating false messages intentionally. This leads to a low value of the resulting TF computed by the CID. Moreover, every NID added to the network must prove its authenticity to CID. NID is only allowed to be a part of IIoT network, once it satisfies the CID by sending authentic messages. Further, if NID is malicious, then the MC would be significantly higher as it will transmit high number of messages and contact every neighbouring node to attract their attention. This results in the minimum values of DDR which affect the network throughput due to the fact that MD jeopardises and consumes most of the network resources.

Further, based on the computed TF, following four scenarios exist for both LID and MD within IIoT network. (1) \textit{H$_{x0}$} signifies the absence of both LID and MD suggesting that neither LID nor MD approached CID to prove their legitimacy. (2) \textit{H$_{x1}$} denotes the case when LID switches from active to the idle channel, hence requiring to prove its legitimacy to the CID. (3) \textit{H$_{x2}$} represents a scenario, where MD tries to replicate LID in order to degrade the network performance, and finally (4) \textit{H$_{x3}$} scenario exist, when both LID and MD prove their legitimacy to the CID as depicted in the Equation \ref{eq:states}.

	\begin{equation}
	\label{eq:states}
	\begin{drcases}
	H_{x0} = \text{neither LID nor MD,}\\
	H_{x1} = \text{LID only,}\\
	H_{x2} = \text{MD only,}\\              
	H_{x3} = \text{LID and MD both.}
	\end{drcases}    
	\end{equation}
	
The presence and absence of MD is indicated by $M^{on}$ and $M^{off}$ respectively. Therefore, the probability of each hypothesis `H$_{sk}$' is denoted by `${\mu}_{k}$' as shown in Equation \ref{eq:states_mu}.  

\begin{equation}
\label{eq:states_mu}
	\begin{drcases}
	\mu_0=Pr(H_{x0})=Pr(H_0),M^{off}=Pr(M^{off}/H_0)Pr_{H0,}\\
	\mu_1=Pr(H_{x1}) = Pr(H_1),M^{off} =Pr(M^{off}/H_1) Pr_{H1,}\\
	\mu_2=Pr(H_{x2}) = Pr(H_0),M^{on}=Pr(M^{on}/H_0) Pr_{H0,}\\
	\mu_3=Pr(H_{x3}) = Pr(H_1),M^{on}=Pr(M^{on}/H_1)Pr_{H1}\\
	\end{drcases}    
	\end{equation}

Furthermore, the attacking strategies are defined for the presence and absence of the IoT device using the attack parameters $\alpha$ and $\beta$ which are determined as: $\alpha$=Pr($M^{on}$/$H_1$) and $\beta$=Pr($M^{on}$/$H_0$). Therefore, previous equation can be extended to equation \ref{eq:states_modify} in the following manner.

	\begin{equation}
	\label{eq:states_modify}
	\begin{drcases}
	\mu_0 = (1-\beta) Pr(H_0),\\
	\mu_1=(1-\alpha) Pr(H_1),\\
	\mu_2=\beta Pr(H_0),\\
	\mu_3=\alpha Pr(H_1).\\
	\end{drcases}    
	\end{equation}

Now let $W_{fa}$ and $W_{m}$ denote the probability of false authentication and non-detection of MD at CID respectively, which are determined as: $W_{fa}$= $Pr(D_{on}/M^{off})$ and $W_{m}$= $Pr(D_{off}/M^{on})$, where $D_{on}$ and $D_{off}$ represents the CID's decision of MD's presence and absence respectively. The system will be in error if CID fails to decide correctly between the presence and absence of MD in the network. Therefore, we define the probability of error $(W_{e})$ to represent this behaviour, which is determined according to the equation \ref{eq:system_error}:

	\begin{equation}
	\label{eq:system_error}
	\begin{split}
	W_e & = Pr(M^{on}, D^{off}) + Pr(M^{off}, D^{on}) \\
	& = Pr(\frac{D^{off}}{M^{on}}) + Pr(\frac{D^{on}}{M^{off}}) Pr(M^{off})\\ 
	& = W_m P(M^{on})+ W_{fa} Pr(M^{off}.)
	\end{split}
	\end{equation}

$SNR_{LID}$ and $SNR_{MD}$ are the the signal-to-noise ratio of LID and MD respectively. Further, to identify the impact of the MD, we have defined attack strength parameter as $\rho$=$SNR_{MD}$/($1+SNR_{LID}$). To detect the compromised IoT devices within the network, we have considered the hypothesis that $H_{x1}$ and $H_{x3}$ will confirm the presence of NID with their respective probabilities $\mu_1$ and $\mu_2$. Therefore, the compromised IoT device $(R_{NID})$ can be identified according to equation \ref{eq:compromised}.

\begin{equation} 
\label{eq:compromised}
	\begin{split}
	R_{NID} & = \mu_1. log_2 (1+ SNR_{LID} + \\
        	&   \mu_3. log_2 (1+ \frac{SNR_{LID}}{1+SNR_{MD}})
	\end{split}
	\end{equation}

\section{Performance Analysis and Simulation Results}
\label{sec:4}
Till now, literature in the field has not projected a probabilistic hypothetical way to compute the trust of IoT devices for analyzing their legitimacy within IIoT networks. Probabilistic hypothetical is the probability to identify the legitimacy of IoT devices based on certain assumed hypothesis as in Equation \ref{eq:states}. In this paper, we have formulated a mathematical model to identify the malicious IoT devices by assuming the probability of each hypothesis. Therefore, in order to evaluate our proposal, the system is validated initially on a small network by constructing a simulation area of 400m $\times$ 400m within MATLAB simulator and inserted 25 IoT devices initially. These devices are identified using unique numbers, which are assigned during the initialization phase in the system.  

Whenever, a NID enters within the respective area, it is required to register and authenticate itself with the CID first. Based on the previous history and monitoring capability of the IoT device, CID either authenticates it or rejects it from the system by assigning different values of TF (either \textit{accept} or \textit{reject}). 

		 
We evaluated our proposed model using three distinct criteria: (1) probability of false authentication versus probability of error, (2) the compromised IoT device versus attack strength ($\alpha$) whose probability is between 0-1,  and (3) and the number of compromised IoT devices versus SNR caused by malicious IoT devices.  Further, important simulation details are highlighted in Table \ref{tab:Simulation_parameters}.

\begin{table}[h]
\caption{Simulation Parameters}
\label{tab:Simulation_parameters}
\centering
\begin{tabular}{ll}
\toprule
\textbf{Parameters}	& \textbf{Details}      	\\
\midrule
Simulation Time		& 80 sec 			\\
Simulated Area		&  400m $\times$ 400m		\\
Total IoT devices (small network)	& (25, 45)		\\
Total IoT devices (large network)	& up to 100 			\\
IoT device Transmission Range	&  120m (Approx.)		\\
Pr(M$^{on}$)		& 	   0.8                 	\\
Pr(M$^{off}$)	    &  0.2                \\
$\beta$             &   0.8                    \\
$\alpha$           &   0.2 \\
\bottomrule
\end{tabular}
\end{table}

\subsection{Simulation Results}
\begin{figure}[h]
 	    	 \centering
  	    	 \includegraphics[width=0.33\textwidth]{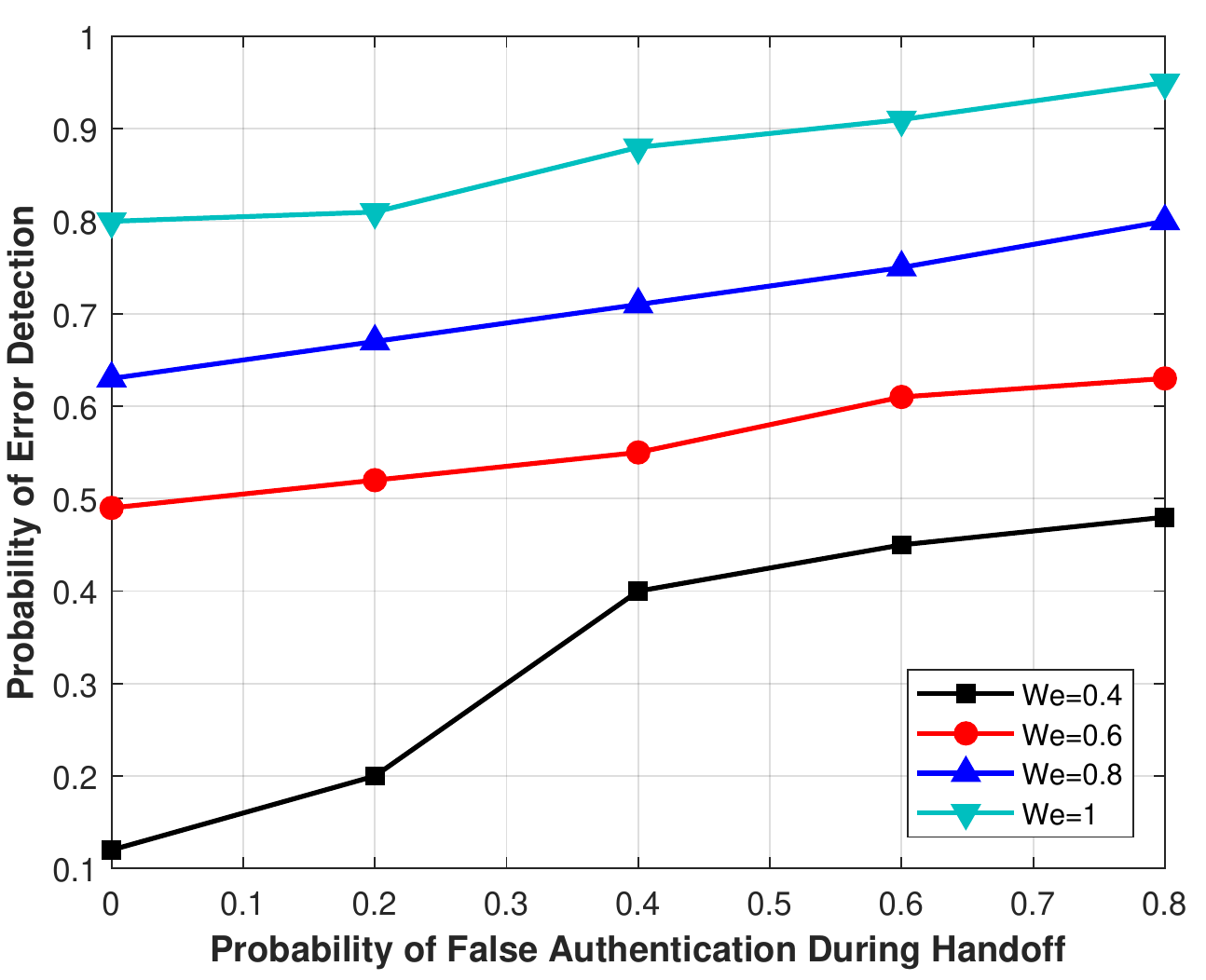}
  	    	 \caption{Probability of false authentication $(W_{fa})$ on the probability of error $(W_e)$}
  	    	 \label{fig:Probablity_of_false_auth_error}
		 \end{figure}

The effect of probability of false authentication $(W_{fa})$ on the probability of error $(W_e)$ for the proposed system is illustrated in the Figure \ref{fig:Probablity_of_false_auth_error}. It is apparent that $W_e$ shows linearly increasing relationship with $W_{fa}$, where the probability of error increases when the authentication probability increases during the hand-off phase. In the ideal situation, the network only contains legitimate IoT devices, implying that $(W_{m})$=1. This shows that the probability of the IoT devices to detect errors is high, as depicted in the Fig. \ref{fig:Probablity_of_false_auth_error}. However, as soon as the network is polluted with MD, the ability of the nodes to detect error decreases. This is due to the fact that the proposed approach identifies the legitimacy of IoT devices by computing their TF. The devices with less TF are considered as MD that would not be involved during the communication process. 

\begin{figure}[ht]
 	    	 \centering
  	    	 \includegraphics[width=0.33\textwidth]{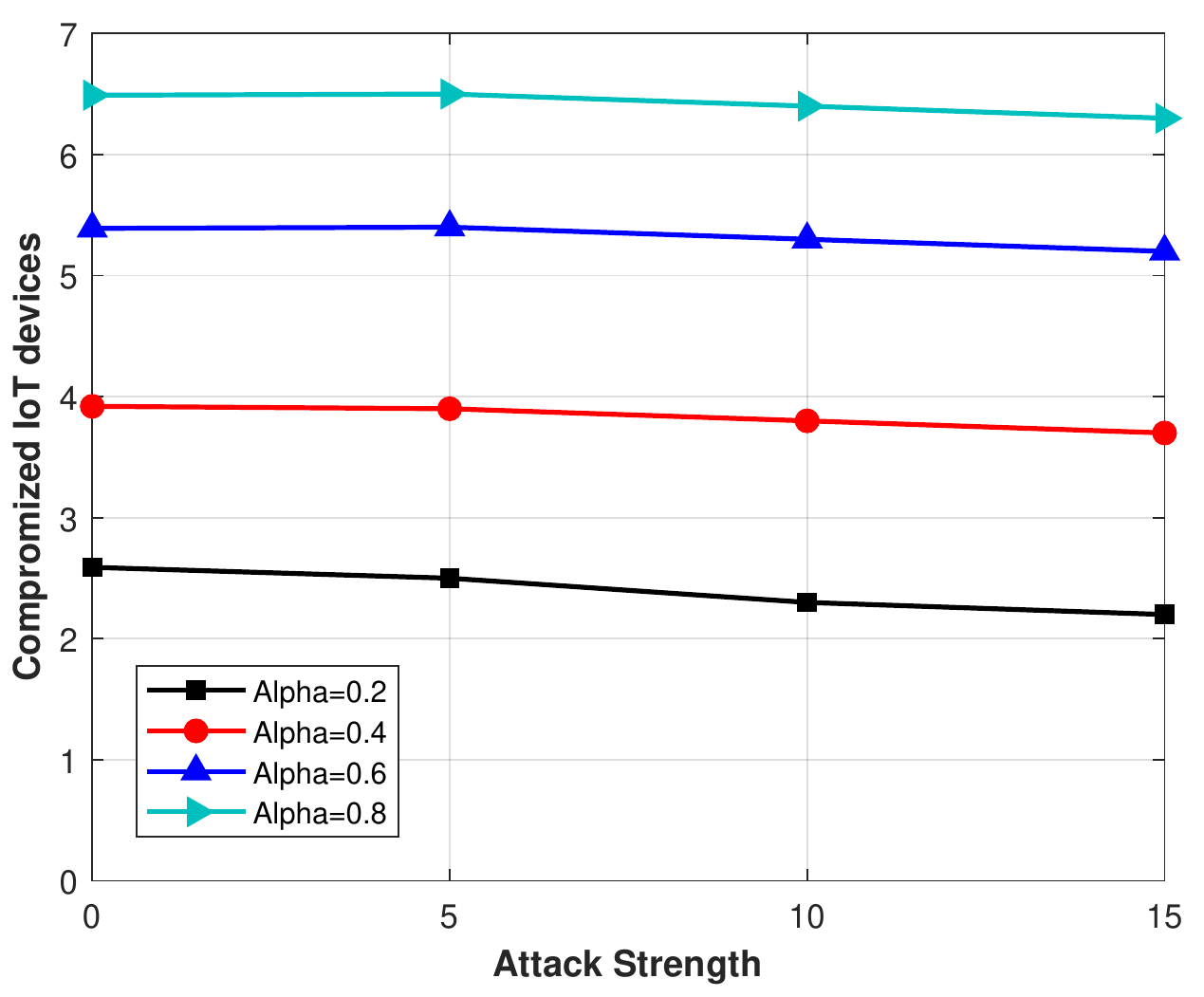}
  	    	 \caption{Compromised IoT device affected by the attack strength}
  	    	 \label{fig:Compromised_IoT_device_affected-by_the_attack_strength}
		 \end{figure}

Fig. \ref{fig:Compromised_IoT_device_affected-by_the_attack_strength} depicts the impact of the attack strength on the newly added IoT devices in the network. It can be seen that for a low attack strength, less number of IoT devices are affected. This compromise increases as soon as the attack strength of the MD increases. However, our proposed solution enables the network to detect MD with high impact due to the fact that the CID only allows a device to be part of the network, if it satisfies the required TF level. Moreover, the small values of $\alpha$ provide enhanced throughput which decreases with increase in the value of $\alpha$.

\begin{figure}[ht]
 	    	 \centering
  	    	 \includegraphics[width=0.33\textwidth]{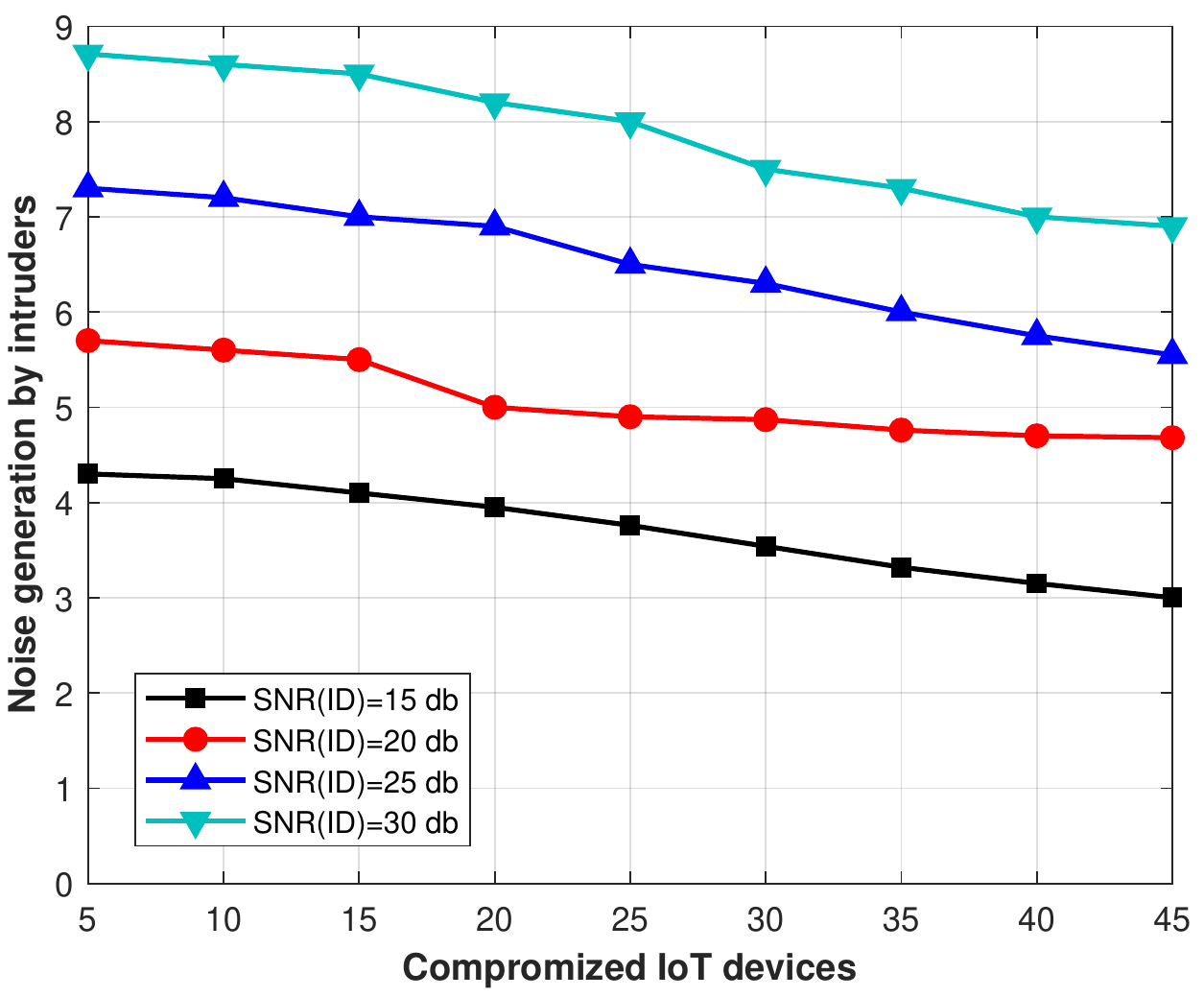}
  	    	 \caption{SNR at compromised IoT receiver due to ID}
  	    	 \label{fig:SNR_Compromised_IoT}
		 \end{figure}

The relationship between the noise generation $(R_{NID})$ and SNR at compromised IoT receiver ($SNR_{MD}$) due to the addition of new IoT devices within the network is presented in Fig. \ref{fig:SNR_Compromised_IoT}. As the presence of the compromised devices increases in the network, the noise generation by intruders at lower SNR is less as compared to higher SNR. The higher the SNR, the higher would be the probability of disruption caused by the intruders in the compromised network. 

As the proposed phenomenon analyzes each and every activity of the devices at each phase such as product manufacturing, storage recording and data delivery, the proposed approach computes the legitimacy of every node before its communication/transmission process.

\subsection{Impact of Blockchain on the IIoT Network}
\begin{figure*}[!tbp]
  \begin{subfigure}[b]{0.325\textwidth}
  \hspace{20mm}
    \includegraphics[width=1\textwidth]{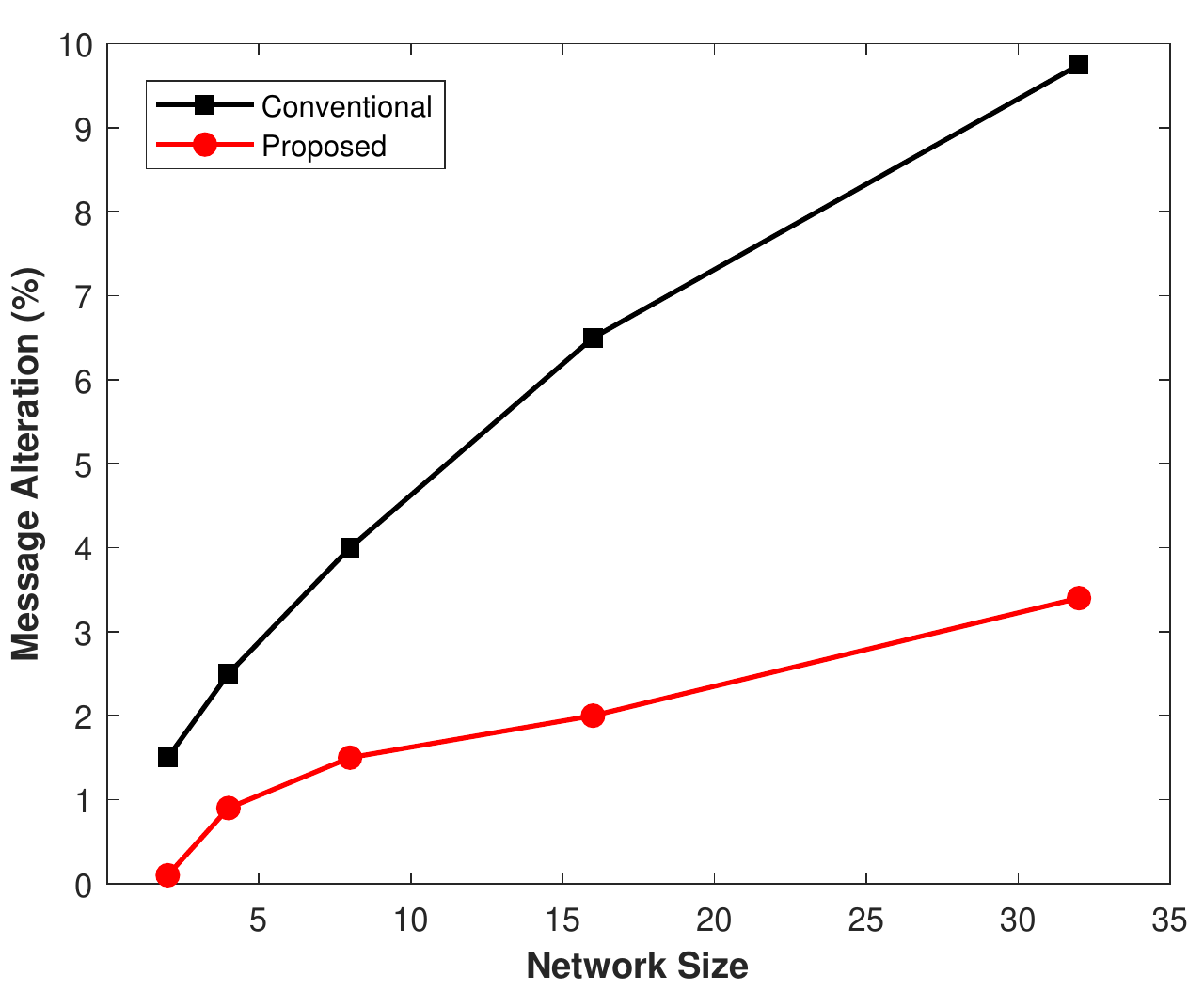}

    \caption{}
    \label{fig:Message_Alteration_01}
  \end{subfigure}
  \hfill
  \begin{subfigure}[b]{0.325\textwidth}
  \hspace{-20mm}
    \includegraphics[width=1\textwidth]{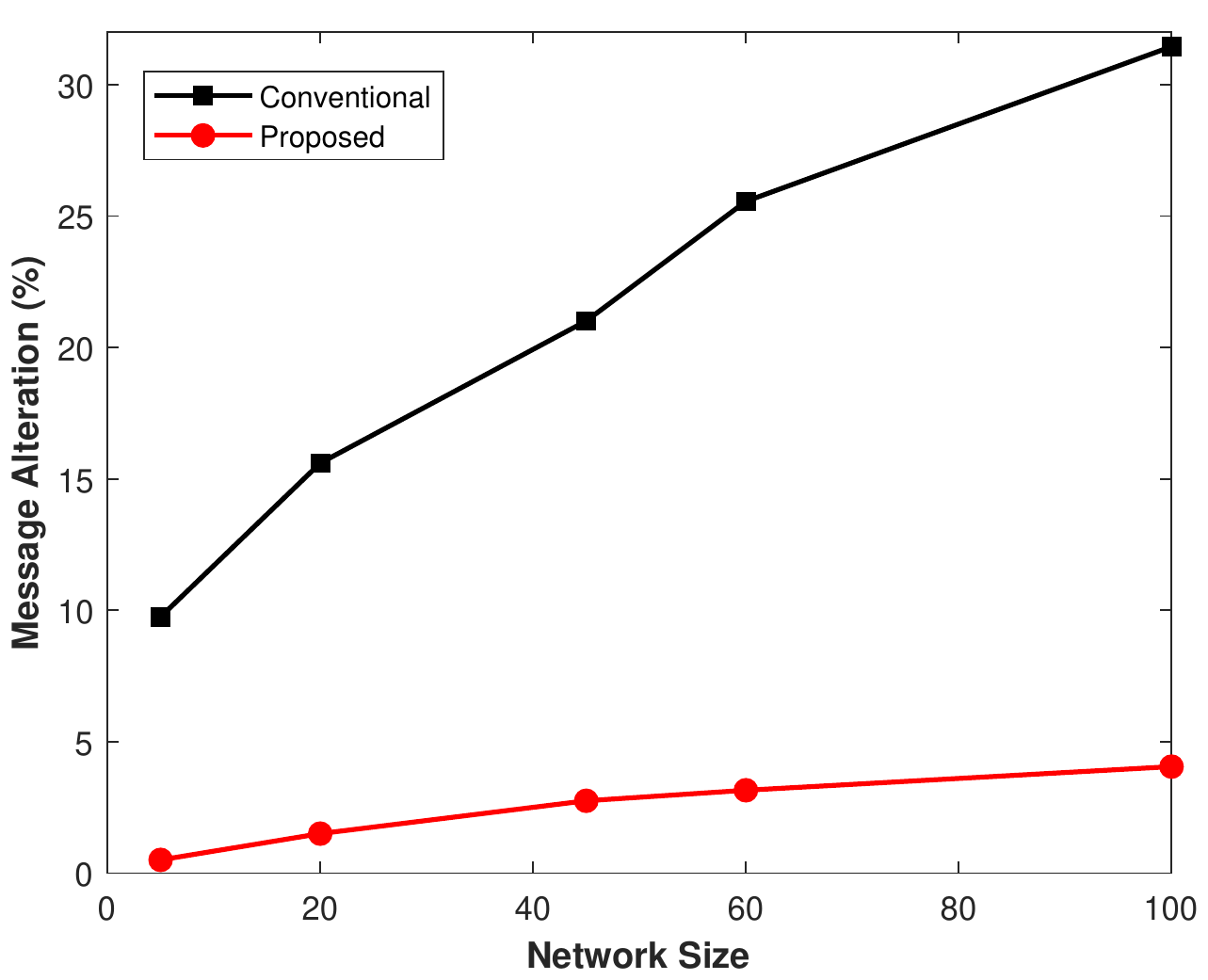}
    \caption{}
    \label{fig:Message_Alteration_02}
  \end{subfigure}
    \vspace{-0.15cm} 
  \caption{Message Alteration for (a) Small Network (b) Large Network}
  \label{fig:Message_alteration_for_different_network_sizes}
\end{figure*}

This paper integrates a private blockchain mechanism \cite{Private_BC1, Private_BC2} at its back-end due to the fact that the data included in IIoT networks is mostly very sensitive. Further, this data must be kept private as the companies need to compete with rival industry. Therefore, the data within IIoT network must be kept private and secure, which can be achieved with the help of private blockchain.
A private blockchain mechanism is the one where accessing and storing records is maintained at different levels in the network and hence it cannot be shared without the consent of administrator (CCU) permission. In this paper, separate blockchain is maintained for different process such as product manufacturing, storing records and delivering products, to name a few. All these blockchains are monitored by CCU and data is modified in the blockchain only with the consent of the administrator. Further, to encourage the cognitive users (CUs) of the network to provide positive data, CCU provides incentives in form of extra credits. 

Further to analyze the integration of blockchain within IIoT networks, we relied on Java where different components (creation, validation and insertion) of blockchain is implemented. Further we validated the proposal against malevolent nodes from security aspects via MATLAB. Initially, 25 nodes are created that may further operate as IoT devices. The blockchain network contains blockchain creation module that is responsible to create blocks along with their previous and current hashes. Further, during the data insertion process of blockchain process, these devices have the ability to add various records in the blockchain. In this paper, we have inserted a single unit data (such as product manufacturing) to analyze it efficiently. 

Finally, the newly added blocks by the miners need to be verified and validated. If the miners are successfully able to verify the block then they are considered as valid and stock is successfully added, otherwise block is rejected. We have evaluated our proposal in presence of intruders for both small and large IIoT networks. A blockchain network of initially 5 ledgers is implemented using Ethereum where each block contains the product information with its respective hash. This ledger grows further as soon as the devices generate data to be stored within the database. Further, we equipped malicious devices in the blockchain network with the ability to alter and delete the recorded data. Simulation results depicted that our proposed solution performs efficiently in presence of these malicious devices.

We have evaluated our proposal in presence of intruders for both small and large IIoT network as depicted in Fig. \ref{fig:Message_alteration_for_different_network_sizes}. A blockchain network of initially 5 ledgers is implemented using Ethereum where each block contains the product information  with its respective hash. This ledger grows further as soon as the devices generate data to be stored within the database. Further, we equipped MDs in the blockchain network with the ability to alter and delete the recorded data as shown in Fig \ref{fig:Compromised_IoT_for_different_network_sizes}. We considered two scenarios, (1) \textit{Conventional scenario}: where no blockchain is considered, and (2) \textit{Proposed scenario} includes blockchain at the back-end. Fig. \ref{fig:Message_alteration_for_different_network_sizes} shows that in the absence of blockchain in the conventional method, the network is affected more as the intruders can alter or delete the data. However, in our proposal, the impact of intrusion is limited as the devices will be unable to delete or alter the data. This is due to the fact that our proposed approach is based on blockchain in the back-end which provides transparency among all the IoT devices and users so that a single change would reflect in all others' database and would become easily traceable.

\subsection{Impact of Trust on the IIoT Network}

\begin{figure*}[!tbp]
  \begin{subfigure}[b]{0.325\textwidth}
  \hspace{20mm}
    \includegraphics[width=1\textwidth]{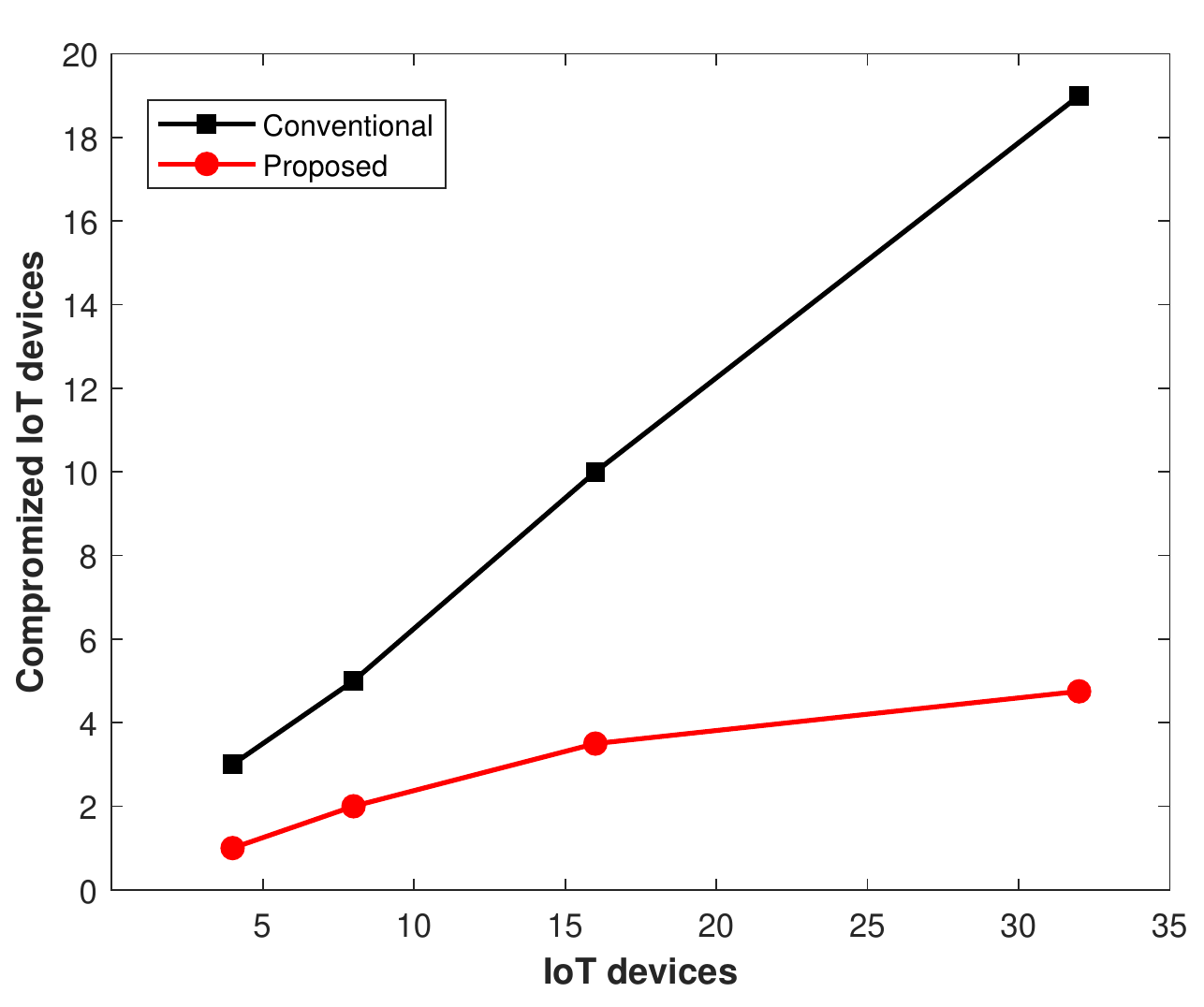}
   
    \caption{}
    \label{fig:Compromised_IoT_Devices_01}
  \end{subfigure}
  \hfill
  \begin{subfigure}[b]{0.325\textwidth}
  \hspace{-20mm}
    \includegraphics[width=1\textwidth]{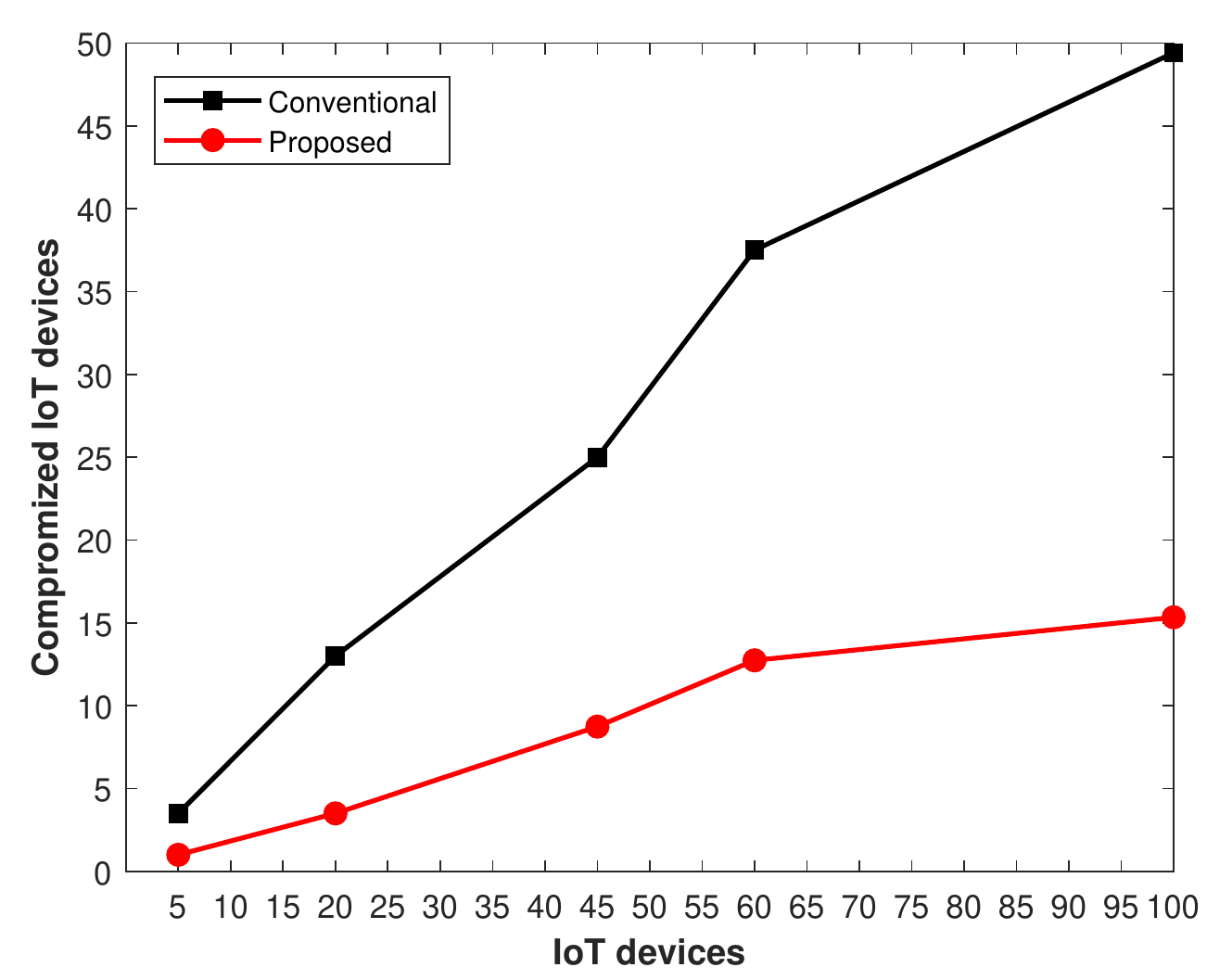}
    \caption{}
    \label{fig:Compromised_IoT_Devices_02}
  \end{subfigure}
       \vspace{-0.15cm} 
  \caption{Compromised IoT deivces for (a) Small Network (b) Large Network}
  \label{fig:Compromised_IoT_for_different_network_sizes}
\end{figure*}

Fig. \ref{fig:Compromised_IoT_for_different_network_sizes} depicts the impact of the compromised IoT devices on the legitimate devices in the network. Conventionally, without applying the trust based mechanism, it becomes very easy to compromise IoT devices by the intruders. However, the current proposal integrates a trust-based mechanism, which only allows the devices to be part of the network after they have been authenticated by the CID. This limits the impact caused by the compromised IoT devices.

However, the proposed phenomenon provides better results as CID computes TF based on their internal behaviour. Similarly, in Figure \ref{fig:Compromised_IoT_for_different_network_sizes}, the proposed phenomenon performs better in terms of compromised miners where during initial establishment of the network, intruders can easily alter the miners, however, miners are selected based on their TF which prevents this scenario. Further, the limitation of this paper is as follows 1) firstly, the proposed scheme is unable to provide a proper verification by comparing the security metrics against any existing method that uses the data sharing process using blockchain mechanism in IIoT networks. None of the authors to the best of our knowledge have worked on datascience mechanisms using blockchain in IIoT while manufacturing or shipping the information, and 2) secondly, the proposed scheme is unable to judge an accurate decision while transmitting the information in real time scenarios as the block verification process may further delay the validation and enhance the changes of other security threats inside the network.


\section{Conclusion}
\label{sec:5}
This paper has proposed a secure framework based on trust management and blockchain to deal with the issues caused by MDs at various levels in IIoT networks. The proposed model identifies the legitimacy of each IoT device by computing its Trust Factor (TF) through an elected Coordinator IoT Device (CID). In order to prevent changes in the information of the local database, a data model based on blockchain is maintained at the back-end to keep track of all the transactions within the industry. The approach is validated extensively for different network sizes and evaluation criteria. Simulation results suggest that our proposed framework achieves 91\% success rate against the network without a blockchain. 


\bibliography{reference.bib}

\begin{thebibliography}{10}

\bibitem{karpatne2017theory}
A.~Karpatne, G.~Atluri, J.~H. Faghmous, M.~Steinbach, A.~Banerjee, A.~Ganguly,
  S.~Shekhar, N.~Samatova, and V.~Kumar, ``Theory-guided data science: A new
  paradigm for scientific discovery from data,'' {\em IEEE Transactions on
  Knowledge and Data Engineering}, vol.~29, no.~10, pp.~2318--2331, 2017.
\newblock doi:10.1109/TKDE.2017.2720168.

\bibitem{Farhan_Commag}
Y.~Mehmood, F.~Ahmad, I.~Yaqoob, A.~Adnane, M.~Imran, and S.~Guizani,
  ``{Internet-of-Things Based Smart Cities: Recent Advances and Challenges},''
  {\em IEEE Communications Magazine}, vol.~55, no.~9, pp.~16--24, 2017.
\newblock doi: 10.1109/MCOM.2017.1600514.

\bibitem{oh2019personal}
H.~Oh, S.~Park, G.~M. Lee, H.~Heo, and J.~K. Choi, ``Personal data trading
  scheme for data brokers in iot data marketplaces,'' {\em IEEE Access},
  vol.~7, pp.~40120--40132, 2019.
\newblock doi:10.1109/ACCESS.2019.2904248.

\bibitem{merolla2014million}
P.~A. Merolla, J.~V. Arthur, R.~Alvarez-Icaza, A.~S. Cassidy, J.~Sawada,
  F.~Akopyan, B.~L. Jackson, N.~Imam, C.~Guo, Y.~Nakamura, {\em et~al.}, ``A
  million spiking-neuron integrated circuit with a scalable communication
  network and interface,'' {\em Science}, vol.~345, no.~6197, pp.~668--673,
  2014.
\newblock doi:10.1126/science.1254642.

\bibitem{wang2014cellular}
C.-X. Wang, F.~Haider, X.~Gao, X.-H. You, Y.~Yang, D.~Yuan, H.~M. Aggoune,
  H.~Haas, S.~Fletcher, and E.~Hepsaydir, ``Cellular architecture and key
  technologies for 5g wireless communication networks,'' {\em IEEE
  communications magazine}, vol.~52, no.~2, pp.~122--130, 2014.
\newblock doi:10.1109/MCOM.2014.6736752.

\bibitem{bertino2017botnets}
E.~Bertino and N.~Islam, ``Botnets and internet of things security,'' {\em
  Computer}, no.~2, pp.~76--79, 2017.
\newblock doi:10.1109/MC.2017.62.

\bibitem{zhou2017computation}
L.~Zhou, D.~Wu, J.~Chen, and Z.~Dong, ``When computation hugs intelligence:
  Content-aware data processing for industrial iot,'' {\em IEEE Internet of
  Things Journal}, vol.~5, no.~3, pp.~1657--1666, 2017.
\newblock doi:10.1109/JIOT.2017.2785624.

\bibitem{huang2019towards}
J.~Huang, L.~Kong, G.~Chen, M.-Y. Wu, X.~Liu, and P.~Zeng, ``Towards secure
  industrial iot: Blockchain system with credit-based consensus mechanism,''
  {\em IEEE Transactions on Industrial Informatics}, 2019.
\newblock doi:10.1109/TII.2019.2903342.

\bibitem{al2018context}
F.~Al-Turjman and S.~Alturjman, ``Context-sensitive access in industrial
  internet of things (iiot) healthcare applications,'' {\em IEEE Transactions
  on Industrial Informatics}, vol.~14, no.~6, pp.~2736--2744, 2018.
\newblock doi:10.1109/TII.2018.2808190.

\bibitem{8621042}
J.~{Wan}, J.~{Li}, M.~{Imran}, and D.~{Li}, ``A blockchain-based solution for
  enhancing security and privacy in smart factory,'' {\em IEEE Transactions on
  Industrial Informatics}, vol.~15, pp.~3652--3660, June 2019.
\newblock doi:10.1109/TII.2019.2894573.

\bibitem{shamsi2018understanding}
J.~A. Shamsi and M.~A. Khojaye, ``Understanding privacy violations in big data
  systems,'' {\em IT Professional}, vol.~20, no.~3, pp.~73--81, 2018.
\newblock doi:10.1109/MITP.2018.032501750.

\bibitem{li2019enhancing}
X.~Li, Q.~Wang, X.~Lan, X.~Chen, N.~Zhang, and D.~Chen, ``Enhancing cloud-based
  iot security through trustworthy cloud service: An integration of security
  and reputation approach,'' {\em IEEE Access}, vol.~7, pp.~9368--9383, 2019.
\newblock doi:10.1109/ACCESS.2018.2890432.

\bibitem{moosavi2016delay}
H.~Moosavi and F.~M. Bui, ``Delay-aware optimization of physical layer security
  in multi-hop wireless body area networks,'' {\em IEEE Transactions on
  Information Forensics and Security}, vol.~11, no.~9, pp.~1928--1939, 2016.
\newblock doi:10.1109/TIFS.2016.2566446.

\bibitem{chen2014collaborative}
Z.~Chen, W.~Dong, H.~Li, P.~Zhang, X.~Chen, and J.~Cao, ``Collaborative network
  security in multi-tenant data center for cloud computing,'' {\em Tsinghua
  Science and Technology}, vol.~19, no.~1, pp.~82--94, 2014.
\newblock doi:10.1109/TST.2014.6733211.

\bibitem{danzi2019delay}
P.~Danzi, A.~E. Kal{\o}r, {\v{C}}.~Stefanovi{\'c}, and P.~Popovski, ``Delay and
  communication tradeoffs for blockchain systems with lightweight iot
  clients,'' {\em IEEE Internet of Things Journal}, vol.~6, no.~2,
  pp.~2354--2365, 2019.
\newblock doi:10.1109/JIOT.2019.2906615.

\bibitem{liu2018evolutionary}
X.~Liu, W.~Wang, D.~Niyato, N.~Zhao, and P.~Wang, ``Evolutionary game for
  mining pool selection in blockchain networks,'' {\em IEEE Wireless
  Communications Letters}, vol.~7, no.~5, pp.~760--763, 2018.
\newblock doi:10.1109/LWC.2018.2820009.

\bibitem{sun2019blockchain}
Y.~Sun, L.~Zhang, G.~Feng, B.~Yang, B.~Cao, and M.~A. Imran,
  ``Blockchain-enabled wireless internet of things: Performance analysis and
  optimal communication node deployment,'' {\em IEEE Internet of Things
  Journal}, 2019.
\newblock doi:10.1109/JIOT.2019.2905743.

\bibitem{yan2017industrial}
J.~Yan, Y.~Meng, L.~Lu, and L.~Li, ``Industrial big data in an industry 4.0
  environment: Challenges, schemes, and applications for predictive
  maintenance,'' {\em IEEE Access}, vol.~5, pp.~23484--23491, 2017.
\newblock doi:10.1109/ACCESS.2017.2765544.

\bibitem{wang2019new}
W.~Wang, L.~Fan, P.~Huang, and H.~Li, ``A new data processing architecture for
  multi-scenario applications in aviation manufacturing,'' {\em IEEE Access},
  vol.~7, pp.~83637--83650, 2019.
\newblock doi:10.1109/ACCESS.2019.2925114.

\bibitem{yu2018high}
S.~Yu, K.~Lv, Z.~Shao, Y.~Guo, J.~Zou, and B.~Zhang, ``A high performance
  blockchain platform for intelligent devices,'' in {\em 2018 1st IEEE
  international conference on hot information-centric networking (HotICN)},
  pp.~260--261, IEEE, 2018.
\newblock doi:10.1109/HOTICN.2018.8606017.

\bibitem{yu2018blockchain}
Y.~Yu, Y.~Li, J.~Tian, and J.~Liu, ``Blockchain-based solutions to security and
  privacy issues in the internet of things,'' {\em IEEE Wireless
  Communications}, vol.~25, no.~6, pp.~12--18, 2018.
\newblock doi:10.1109/MWC.2017.1800116.

\bibitem{oh2020competitive}
H.~Oh, S.~Park, G.~M. Lee, J.~K. Choi, and S.~Noh, ``Competitive data trading
  model with privacy valuation for multiple stakeholders in iot data markets,''
  {\em IEEE Internet of Things Journal}, vol.~7, no.~4, pp.~3623--3639, 2020.

\bibitem{hasan2022trustworthy}
H.~R. Hasan, K.~Salah, I.~Yaqoob, R.~Jayaraman, S.~Pesic, and M.~Omar,
  ``Trustworthy iot data streaming using blockchain and ipfs,'' {\em IEEE
  Access}, vol.~10, pp.~17707--17721, 2022.

\bibitem{lam2022dynamical}
A.~N. Lam, {\O}.~Haugen, and J.~Delsing, ``Dynamical orchestration and
  configuration services in industrial iot systems: An autonomic approach,''
  {\em IEEE Open Journal of the Industrial Electronics Society}, vol.~3,
  pp.~128--145, 2022.

\bibitem{Private_BC1}
S.~{Zhao}, S.~{Li}, and Y.~{Yao}, ``Blockchain enabled industrial internet of
  things technology,'' {\em IEEE Transactions on Computational Social Systems},
  vol.~6, pp.~1442--1453, Dec 2019.
\newblock doi:10.1109/TCSS.2019.2924054.

\bibitem{Private_BC2}
D.~{Liu}, A.~{Alahmadi}, J.~{Ni}, X.~{Lin}, and X.~{Shen}, ``Anonymous
  reputation system for iiot-enabled retail marketing atop pos blockchain,''
  {\em IEEE Transactions on Industrial Informatics}, vol.~15, pp.~3527--3537,
  June 2019.
\newblock doi:10.1109/TII.2019.2898900.

\end{thebibliography}
\bibliographystyle{ieeetr}

\end{document}